\documentclass[aps,pra,twocolumn,floatfix]{revtex4}
\usepackage{bm}
\usepackage{epsfig}

\begin{document}
\title{Bohr's 1913 molecular model revisited}
\author{Anatoly A. Svidzinsky$^{\ast,\dagger}$,
Marlan O. Scully$^{\ast,\dagger,\ddagger}$ and Dudley R.
Herschbach$^\S$}
\affiliation{
$^\ast$Depts. of Chemistry, and Mechanical and Aerospace Engineering,
Princeton University, Princeton, NJ 08544\\
$^\dagger$Depts. of Physics, Chemical and Electrical Engineering,
Texas A\&M University, TX 77843-4242 \\
$^\ddagger$Max-Planck-Institut f\"ur Quantenoptik, D-85748 Garching,
Germany\\
$^\S$Department of Chemistry and Chemical Biology,
Harvard University, Cambridge, MA 02138
}

\date{\today }

\begin{abstract}
It is generally believed that the old quantum theory, as presented by Niels
Bohr in 1913, fails when applied to few electron systems, such as the H$_2$
molecule. Here we find new solutions within the Bohr theory that describe
the potential energy curve for the lowest singlet and triplet states of H$_2
$ about as well as the early wave mechanical treatment of Heitler and
London. We also develop a new interpolation scheme which substantially
improves the agreement with the exact ground state potential curve of H$_2$
and provides a good description of more complicated molecules such as LiH, Li%
$_2$, BeH and He$_2$.
\end{abstract}

{\bf Classification:} PHYSICAL SCIENCES: Physics, Chemistry

\maketitle

The Bohr model \cite{Bohr1} for a one-electron atom played a major historical
role and still offers pedagogical appeal. However, when applied to the
simple H$_2$ molecule, the ``old quantum theory'' proved unsatisfactory \cite
{Vlec22,Somm23}. Here we show that a simple extension of the original Bohr
model describes the potential energy curves $E(R)$ for the lowest singlet
and triplet states about as well as the first wave mechanical treatment due
to Heitler and London \cite{Heit27}.

We find the Bohr model of H$_2$ admits other solutions than the
symmetric one he considered
(pictured in Fig. \ref{draw}) \cite{refer}.
These provide a fairly good description of the ground state $%
E(R)$ (curve 2 in Fig. \ref{f2c}) at large as well as small internuclear
spacing $R$, in contrast with the result of Bohr (curve 1 in Fig. \ref{f2c}).

Clearly the Bohr picture of a molecule goes wrong at large $R$.
Any realistic model must show the ground state potential energy function
dissociating to H+H. Sommerfeld, in his seminal book \cite{Somm23}, provided
an apt assessment: ``We shall now describe a little more fully the model
that Bohr has suggested for the constitution of the hydrogen molecule H$_2$,
although, nowadays, we can take only a historical interest in it.'' After
some discussion he asks: ``But is it correct?'' To which he answers: ``Only
a short while ago, even while this book was in its first edition, we were
inclined to accept it''. And later he concludes: ``Thus the true model of
the H$_2$ molecule is still unknown. It will hardly be as symmetrically
built as the model exhibited in Fig. 22''. His Fig. 22 is the same as our
symmetric configuration in Fig. \ref{f2c}.

It is somewhat ironic that the Bohr picture of the molecule never caught on.
As with the Bohr atomic picture, it contains valuable insight, and can
provide a good analytical description of molecular behavior. Sommerfeld even
sensed that the symmetric configuration was suspect. In Fig.
\ref{f2c} we present a
simple continuation of the line of thought that Bohr was following which is
indeed asymmetric and provides a good quantitative picture of H$_2$ at small
and large $R$. We next outline Bohr's insightful picture and our extensions.

Figure \ref{f1c} displays the Bohr model for a hydrogen molecule, in which
two nuclei with charges $Z|e|$ are separated by a fixed distance $R$
(adiabatic approximation) and the two electrons move in the space between
them. The model assumes that the electrons move with constant speed on
circular trajectories of radii $\rho _1=\rho _2=\rho $. The circle centers
lie on the molecule axis $z$ at the coordinates $z_1=\pm z_2=z$. The
separation between the electrons is constant. The net force on each electron
consists of three contributions: attractive interaction between an electron
and the two nuclei, the Coulomb repulsion between electrons, and the
centrifugal force on the electron. We proceed by writing the energy function
$E=T+V$, where the kinetic energy $T=p_1^2/2m+p_2^2/2m$ for electrons 1 and
2 can be obtained from the quantization condition that the circumference is
equal to the integer number $n$ of the electron de Broglie wavelengths $2\pi
\rho =nh/p$, so that we have $T=p^2/2m=n^2\hbar ^2/2m\rho ^2$. All distances
we express in terms the Bohr length $a_0=\hbar ^2/me^2$, where $m$ is the
electron mass, and take $e^2/a_0$ as a unit of energy. The Coulomb potential
energy $V$ is given by
\begin{equation}
\label{b1}V=-\frac Z{r_{a1}}-\frac Z{r_{b1}}-\frac Z{r_{a2}}-\frac Z{r_{b2}}+%
\frac 1{r_{12}}+\frac{Z^2}R,
\end{equation}
where $r_{ai}$ ($i=1,2$) and $r_{bi}$ are the distances of the $i$th
electron from nuclei A and B, as shown in Fig. \ref{f1c} (bottom), $r_{12}$
is the separation between electrons. In cylindrical coordinates the
distances are%
$$
r_{ai}=\sqrt{\rho _i^2+\left( z_i-\frac R2\right) ^2},\quad r_{bi}=\sqrt{%
\rho _i^2+\left( z_i+\frac R2\right) ^2},
$$
$$
r_{12}=\sqrt{(z_1-z_2)^2+\rho _1^2+\rho _2^2-2\rho _1\rho _2\cos \phi },
$$
here $R$ is the internuclear spacing and $\phi $ is the dihedral angle
between the planes containing the electrons and the internuclear axis. The
Bohr model energy for a homonuclear molecule having charge $Z$ is then given
by (here we discuss the case $n=1$)
\begin{equation}
\label{b2}E=\frac 12\left( \frac 1{\rho _1^2}+\frac 1{\rho _2^2}\right)
+V(\rho _1,\rho _2,z_1,z_2,\phi ,R).
\end{equation}
Possible electron configurations correspond to extrema of Eq. (\ref{b2}).
There are four such configurations for which $\rho _1=\rho _2=\rho $, $%
z_1=\pm z_2=z$ and $\phi =\pi $, $0$; they are pictured in Fig. \ref{f2c}
(upper panel).

\begin{widetext}

\vspace{2cm}

\begin{figure}[t]
\bigskip
\centerline{\epsfxsize=0.75\textwidth\epsfysize=0.55\textwidth
\epsfbox{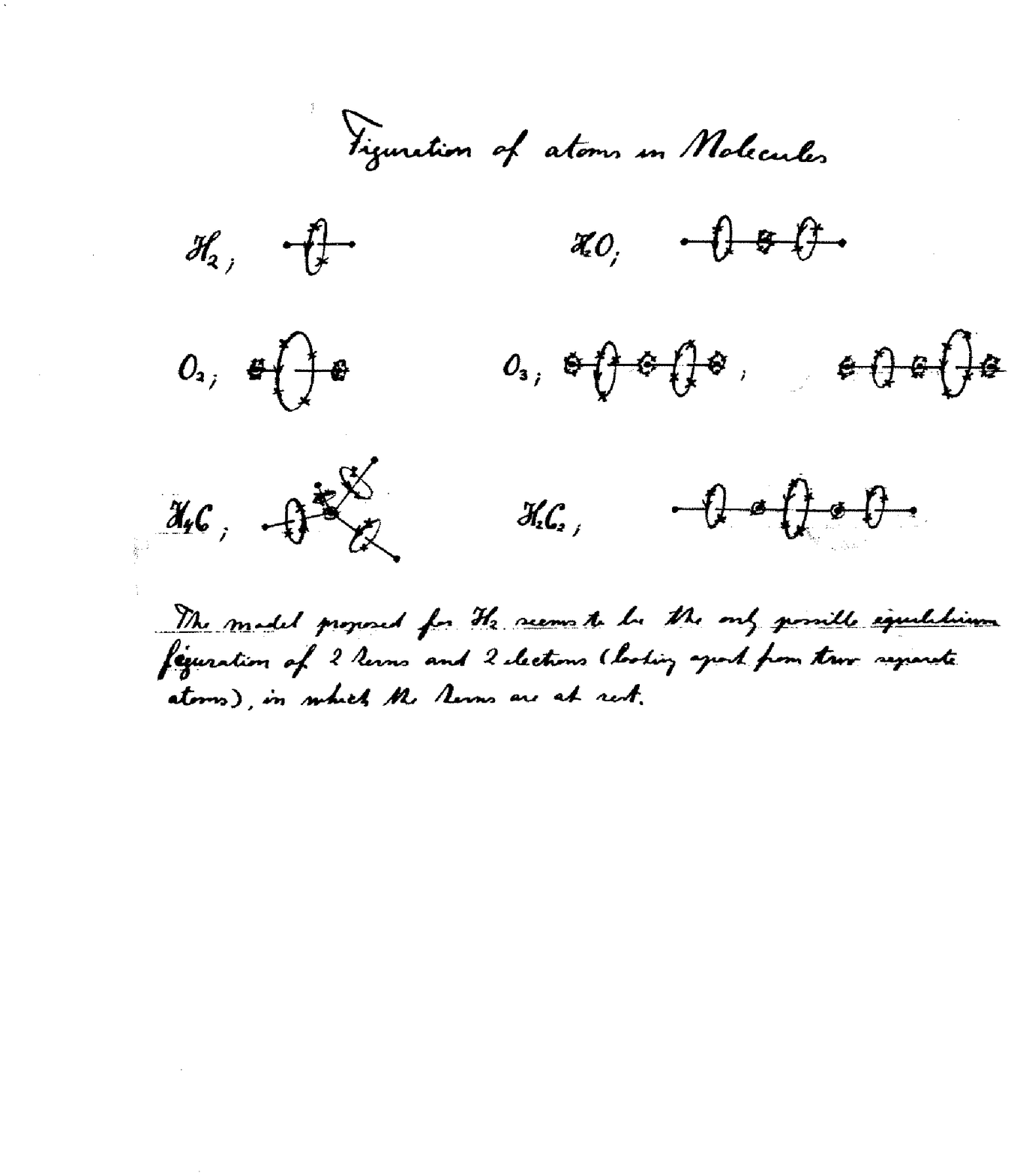}}




\caption{Molecular configurations as sketched by
Niels Bohr; from an unpublished manuscript \cite{Bohr85}, intended as an
appendix to his 1913 papers.}

\label{draw}
\end{figure}

\end{widetext}

\begin{figure}[h]
\bigskip
\centerline{\epsfxsize=0.35\textwidth\epsfysize=0.35\textwidth
\epsfbox{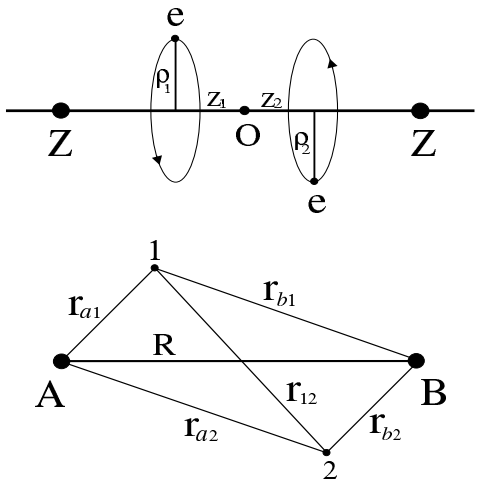}}

\caption{Cylindrical coordinates (top) and electronic distances (bottom) in
H$_2$ molecule. The nuclei $Z$ are fixed at a distance $R$ apart.
The two
electrons rotate about the internuclear axis $z$ with coordinates $\rho_1$,
$z_1$ and $\rho_2$, $z_2$ respectively; the dihedral angle $\phi$ between the
($\rho_1,z_1$) and ($\rho_2,z_2$) planes remains constant at either $\phi=\pi$
or $\phi=0$. The sketch corresponds to configuration
2 of Fig. \ref{f2c}, with $\phi=\pi$.
}
\label{f1c}
\end{figure}

In Fig. \ref{f2c} (lower panel) we plot $E(R)$ for the four Bohr model
configurations (solid curves), together with ``exact'' results (dots)
obtained from extensive variational wave mechanical calculations for the
singlet ground state $^1\Sigma _g^{+}$, and the lowest triplet state, $%
^3\Sigma _u^{+}$ \cite{Scha84}. In the model, the three configurations 1, 2,
3 with the electrons on opposite sides of the internuclear axis ($\phi =\pi $%
) are seen to correspond to singlet states, whereas the other solution 4
with the electrons on the same side ($\phi =0$) corresponds to the triplet
state. At small internuclear distances, the symmetric configuration 1
originally considered by Bohr agrees well with the ``exact'' ground state
quantum energy; at larger $R$, however, this configuration climbs far above
the ground state and ultimately dissociates to the doubly ionized limit, 2H$%
^{+}$+2e. In contrast, the solution for the asymmetric configuration 2
appears only for $R>1.20$ and in the large $R$ limit dissociates to two H
atoms. The solution for asymmetric configuration 3 exists only for $R>1.68$
and climbs steeply to dissociate to an ion pair, H$^{+}$+H$^{-}$. The
asymmetric solution 4 exists for all $R$ and corresponds throughout to
repulsive interaction of two H atoms.

\begin{figure}
\bigskip
\centerline{\epsfxsize=0.5\textwidth\epsfysize=0.68\textwidth
\epsfbox{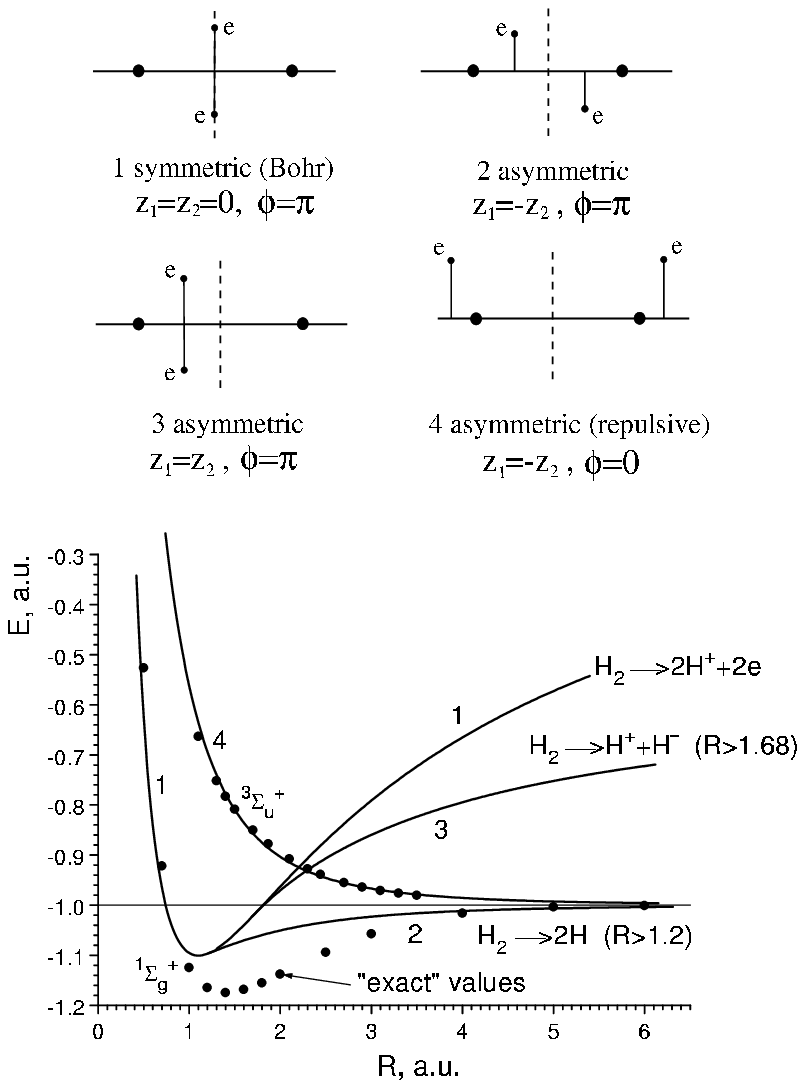}}

\caption{
Energy $E(R)$ of H$_2$ molecule
for four electron configurations (top) as a
function of internuclear distance $R$ calculated within the Bohr model
(solid lines) and the ``exact'' ground $^1\Sigma_g^+$
and first excited $^3\Sigma_u^+$
state energy (dots) \cite{dot}. Unit
of energy is 1 a.u.$=27.21$ eV, and
unit of distance is the Bohr radius.
Please note a similarity between the symmetric
configuration 1 and Bohr's sketch of H$_2$ molecule in Fig. \ref{draw}.
}
\label{f2c}
\end{figure}

The simplistic Bohr model provides surprisingly accurate energies for the
ground singlet state at large and small internuclear distances and for the
triplet state over the full range of $R$. Also, the model predicts the
ground state is bound with an equilibrium separation $R_e\approx 1.10$ and
gives the binding energy as $E_B\approx 0.100$ a.u.$=2.73$ eV. The
Heitler-London calculation, obtained from a two-term variational function,
obtained $R_e=1.51$ and $E_B=3.14$ eV \cite{Heit27}, whereas the ``exact''
results are $R_e=1.401$ and $E_B=4.745$ eV \cite{Scha84,Parr}. For the
triplet state, as seen in Fig. \ref{f2c}, the Bohr model gives remarkably
close agreement with the ``exact'' potential curve and is in fact much
better than the Heitler-London result (which, e.g., is 30\% high at $R=2$).
One should mention that in 1913, Bohr found only the symmetric configuration
solution, which fails drastically to describe the ground state dissociation
limit. Although a variety of modifications were later considered \cite
{Lang21,Vlec22}, to our knowledge the other three solutions of the simplest
model have never been discussed in the literature. One should certainly pay
tribute to Bohr's planetary model proposed long before the development of
quantum mechanics. It is somewhat ironic that the Bohr model can be derived
from quantum mechanics in the limit of large dimensions \cite{Svid05}.

We conclude this first portion of our paper with a quick sketch of the way
the calculations are carried out in order to emphasize how simple the
present analysis is, as compared to the many particle Schr\"odinger
equation. For example, for the configuration 2, with $z_1=-z_2=z$, $\phi
=\pi $, the extremum equations $\partial E/\partial z=0$ and $\partial
E/\partial \rho =0$ read%
$$
\frac{Z(R/2-z)}{\left[ \rho ^2+(R/2-z)^2\right] ^{3/2}}+\frac z{4[\rho
^2+z^2]^{3/2}}-
$$
\begin{equation}
\label{b3}\frac{Z(R/2+z)}{\left[ \rho ^2+(R/2+z)^2\right] ^{3/2}}=0,
\end{equation}
$$
\frac{Z\rho }{\left[ \rho ^2+(R/2-z)^2\right] ^{3/2}}+\frac{Z\rho }{\left[
\rho ^2+(R/2+z)^2\right] ^{3/2}}-
$$
\begin{equation}
\label{b4}\frac \rho {4[\rho ^2+z^2]^{3/2}}=\frac 1{\rho ^3},
\end{equation}
which are seen to be equivalent to Newton's second law applied to the motion
of each electron. Eq. (\ref{b3}) specifies that the total Coulomb force on
the electron along the $z-$axis is equal to zero; Eq. (\ref{b4}) specifies
that the projection of the Coulomb force toward the molecular axis equals
the centrifugal force. At any fixed internuclear distance $R$, these
algebraic equations determine the constant values of $\rho $ and $z$ that
describe the electron trajectories. Substituting these values back into Eq. (%
\ref{b2}) yields $E(R)$. Similar force equations describe the other extremum
configurations.

The simple Bohr model is also useful in describing more complicated diatomic
molecules. For $N$ electrons the model reduces to finding configurations
that deliver extrema of the energy
\begin{equation}
\label{b5}E=\frac 12\sum_{i=1}^N\frac{n_i^2}{\rho _i^2}+V({\bf r}_1,{\bf r}%
_2,...,{\bf r}_N,R),
\end{equation}
In such a formulation of the model there is no need to specify electron
trajectories and also incorporate nonstationary electron motion. One can obtain
the energy function (\ref{b5}) from dimensional scaling analysis of the
Schr\"odinger equation in large-D limit \cite{Svid05}. This provides a link
between the old (Bohr-Sommerfeld) and the new (Heisenberg-Schr\"odinger)
quantum mechanics.

\begin{figure}
\bigskip
\centerline{\epsfxsize=0.4\textwidth\epsfysize=0.3\textwidth
\epsfbox{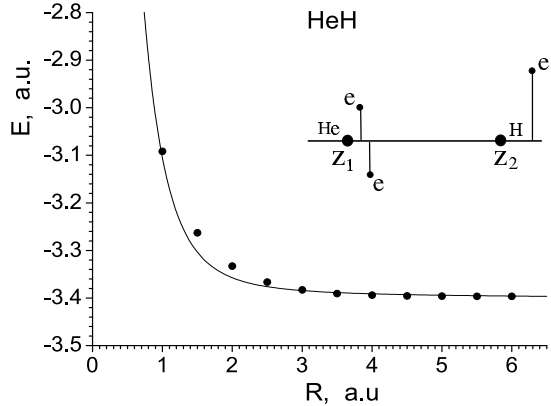}}

\caption{Energy $E(R)$ of HeH molecule
for the shown electron configuration calculated within the Bohr model
for $n_1=n_2=n_3=1$, $Z_{He}^{\text{eff}}=1.954$ (solid line) and the ``exact''
ground state energy (dots).  }
\label{hehcc}
\end{figure}

Next we discuss the ground state potential curve of HeH. To incorporate the
Pauli exclusion principle one can use a prescription based on the sequential
filling of the electron levels. In the case of HeH the three electrons
cannot occupy the same lowest level of HeH$^{++}$. Therefore, for the
configuration with $n_1=n_2=n_3=1$, the true ground state energy corresponds
to a saddle point rather than to a global minimum. Such a configuration is
pictured in Fig. \ref{hehcc} (insert). In order to obtain the correct
dissociation limit we assign the helium nucleus an effective charge $Z_{%
\text{He}}^{\text{eff}}=1.954$. Fig. \ref{hehcc} shows the ground state
potential curve of HeH in the Bohr model (solid curve) and the ``exact''
result (dots) obtained from extensive variational wave mechanical
calculations \cite{dot}. The Bohr model gives remarkably close agreement
with the ``exact'' potential energy curve.

\begin{figure}
\bigskip
\centerline{\epsfxsize=0.4\textwidth\epsfysize=0.4\textwidth
\epsfbox{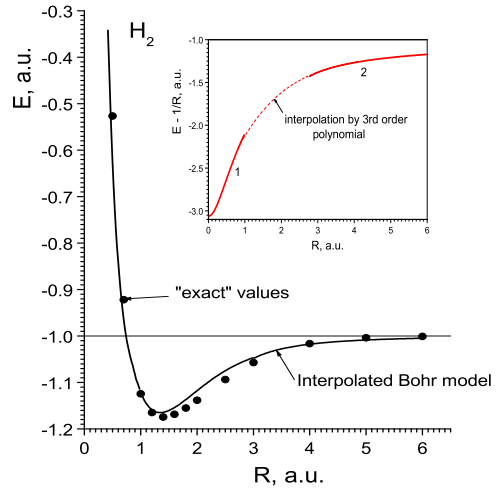}}

\caption{
Ground state $E(R)$ of H$_2$ molecule calculated within the interpotated Bohr
model (solid line) and the
``exact'' energy (dots) \cite{dot}.
Insert shows $E(R)$ with no $1/R$ term. Curves 1
and 2 are obtained based on the quantization relative to the molecular axis
(small $R$) and the nearest nuclei (large $R$) respectively. Dashed line is the
interpolation between two regions.
}
\label{h2int}
\end{figure}

We have found a simple means to improve significantly the
Bohr model results for bound electronic states.  The original model
assumes quantization of the electron angular momentum relative to the
molecular axis.  As seen in Fig. \ref{f2c}, this yields a quite accurate
description of the H$_2$ ground state $E(R)$ at small $R$, but becomes less
accurate at larger internuclear separation.  An improvement emerges
from the following observation.  At large $R$ each electron in H$_2$ feels
only the nearest nuclear charge.  Accordingly, as $R\rightarrow \infty $, we  have two weakly interacting, neutral H atoms.
Therefore, at large $R$ quantization of the momentum relative to the
nearest nuclei, rather than to the molecular axis yields a better
description of the physics. This
leads to the following expression for the energy of the $H_2$ molecule
\begin{equation}
\label{b16}E=\frac 12\left( \frac{n_1^2}{r_{a1}^2}+\frac{n_2^2}{r_{b2}^2}%
\right) +V(r_{a1},r_{b1},r_{a2},r_{b2},r_{12},R).
\end{equation}
For $n_1=n_2=1$ and $R>2.8$ the expression (\ref{b16}) has a local minimum
for the asymmetric configuration 2 of Fig. \ref{f2c}. We plot the
corresponding $E(R)$ without the $1/R$ term in the insert of Fig. \ref{h2int}
(curve 2). At $R<2.8$ the local minimum disappears and electrons collapse
into the opposite nuclei. At small $R$ we apply the quantization condition
relative to the molecular axis which yields the curve 1 in Fig. \ref{h2int}.
To find $E(R)$ at intermediate separation we smoothly connect the two
regions by a third order polynomial (dashed line). Addition of the $1/R$
term yields the final potential curve, plotted in Fig. \ref{h2int}. This
simple interpolated Bohr model provides good agreement with the ``exact''
potential curve over the full range of $R$.

\begin{figure}
\bigskip
\centerline{\epsfxsize=0.5\textwidth\epsfysize=0.4\textwidth
\epsfbox{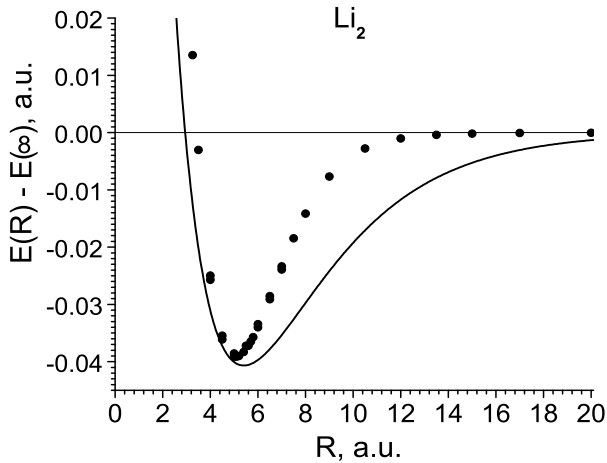}}

\caption{Ground state energy $E(R)$
of Li$2$ molecule calculated within the interpolated Bohr
model (solid line) and the
``exact'' energy (dots).
}
\label{Li2}
\end{figure}

Next we consider the Li$_2$ molecule. If we neglect inner shell electrons of
Li then the Li$_2$ molecule becomes similar to an excited state of H$_2$
with $n_1=n_2=n=2$ in Eq. (\ref{b16}). Rescaling coordinates in Eq. (\ref
{b16}) as $r\rightarrow n^2r$, $R\rightarrow n^2R$ yields the energy
function
\begin{equation}
\label{b17}E=\frac 1{n^2}\left\{ \frac 1{2r_{a1}^2}+\frac 1{2r_{b2}^2}%
+V(r_{a1},r_{b1},r_{a2},r_{b2},r_{12},R)\right\} .
\end{equation}
Hence, the ground state potential curve of Li$_2$ can be obtained from the
ground state $E(R)$ of H$_2$ using the following relation
\begin{equation}
\label{b18}E_{\text{Li}_2}(R)-E_{\text{Li}_2}(\infty )=\frac 1{n^2}\left[ E_{%
\text{H}_2}(n^2R)-E_{\text{H}_2}(\infty )\right] .
\end{equation}
The result is shown in Fig. \ref{Li2} (solid line). For Li$_2$ the Bohr model
gives the binding energy $E_B=1.10$ eV which is very close to the
``exact'' value of $E_B=1.05$ eV.

\begin{figure}
\bigskip
\centerline{\epsfxsize=0.5\textwidth\epsfysize=0.45\textwidth
\epsfbox{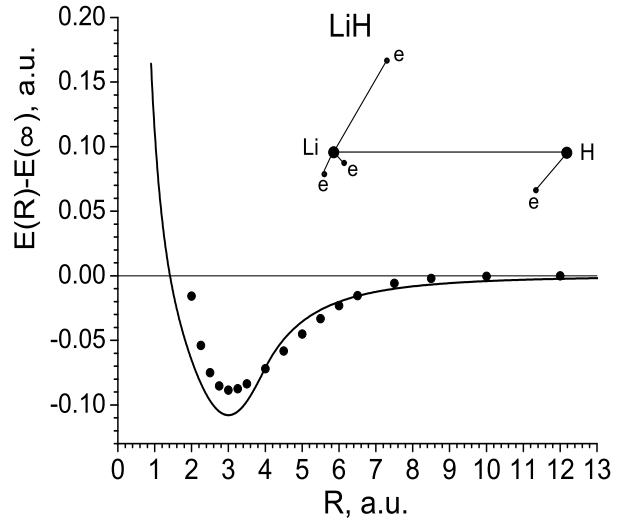}}

\caption{Electron configuration and the
ground state energy $E(R)$ of LiH molecule as a
function of internuclear distance $R$ calculated within the interpotated Bohr
model (solid line) and the
``exact'' energy (dots). }
\label{Lih}
\end{figure}

\begin{figure}
\bigskip
\includegraphics[angle=270,width=8cm]{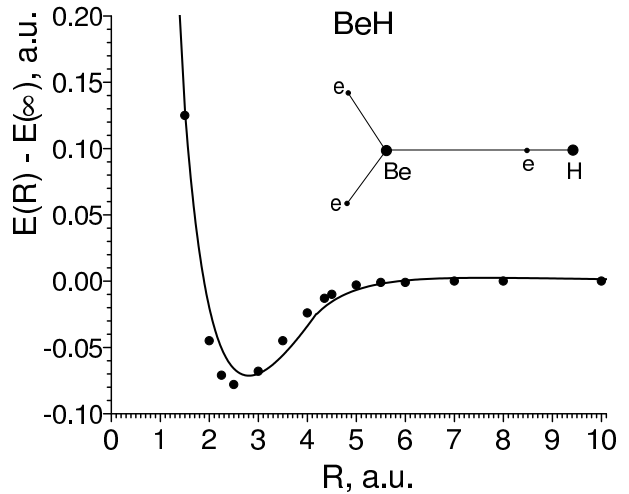}


\caption{Ground state energy $E(R)$
of BeH molecule calculated within the interpotated Bohr
model (solid line) and the
``exact'' energy (dots). Insert shows the electron configuration at large $R$;
only outer shell Be electrons are displayed.
}
\label{BeH}
\end{figure}

As an example of application of the extended Bohr model to other diatomic
molecules, we discuss the ground state $E(R)$ of LiH. The Li atom contains
three electrons two of which fill the inner shell. Only the outer electron
with the principal quantum number $n=2$ is important in formation of the
molecular bond. Applying a similar approach to that used to obtain Fig. \ref
{h2int}, we find $E(R)$ for LiH as shown in Fig. \ref{Lih} (solid line), while
dots are the ``exact'' numerical answer.
This simple
extension of the Bohr model provides a good quantitative description
of the LiH potential curve.  In this treatment, the essential
difference from H$_2$ arises simply because in LiH the $n = 2$ electron
from Li is much more weakly bound than the $n = 1$ electron from H,
with the result that for LiH the binding energy is twofold less than
for H$_2$ and the equilibrium separation roughly twice as large.  As
seen in Fig. \ref{BeH}, the same procedure also gives a good potential curve
for BeH, a relatively complex five electron system.

\begin{figure}
\bigskip
\centerline{\epsfxsize=0.5\textwidth\epsfysize=0.45\textwidth
\epsfbox{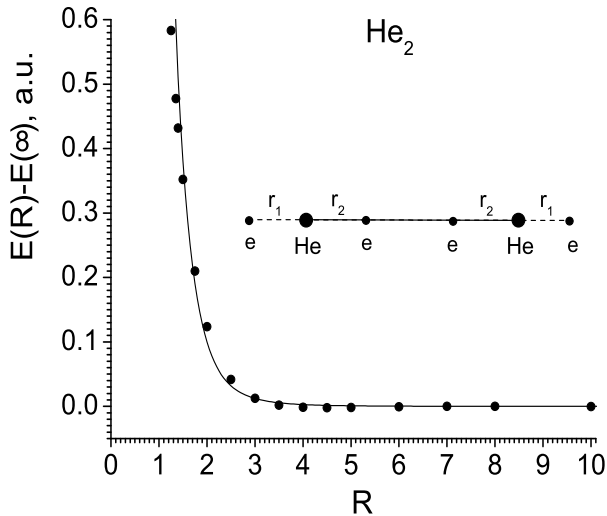}}

\caption{Ground state energy $E(R)$
of He$_2$ molecule calculated within the Bohr
model (solid line) and the
``exact'' energy (dots). Insert shows the electron configuration.
}
\label{He2}
\end{figure}

Finally we show how our very simple analysis yields very accurate potential
curve for the He$_2$ molecule.
We apply the Bohr model with momentum quantization
relative to the nearest nuclei and assume the electron configuration as
shown in the insert of Fig. \ref{He2}. Then the problem reduces to finding
minimum of the following energy function
$$
E=\frac 1{r_1^2}+\frac 1{r_2^2}-\frac{2Z}{r_1}-\frac{2Z}{r_2}+\frac 2{r_1+r_2%
}+\frac 2{R+r_1-r_2}+
$$
\begin{equation}
\label{b19}\frac 1{R+2r_1}+\frac 1{R-2r_2}+\frac{Z^2}R.
\end{equation}
Minimization of this simple expression at fixed $R$ leads the potential
energy curve pictured in Fig. \ref{He2} (solid line). The curve essentially
passes through the ``exact'' dots over the full range of $R$.

In conclusion, we find a simple extension of the Bohr molecular model which
gives a clear physical picture of how electrons create chemical bonding. At
the same time, the description is surprisingly accurate providing good
potential energy curves for relatively complex many body systems.

We wish to thank M. Kim, S. Chin, and G. S\"ussmann for helpful discussions.
This work was supported by the Robert A. Welch Foundation Grant A-1261, ONR,
AFOSR, DARPA and NSF Grant CHE-9986027 (D.R.H).


\begin{thebibliography}{99}
\bibitem{refer}  In response to a referee who said ``[I]t is hard to believe
that Bohr and his students were not aware of the other solutions'', we have
researched the issue further. In fact Bohr says in his hand written notes
\cite{Bohr85}:
``The model proposed for H$_2$ seems to be
the only possible equilibrium figuration of 2 kerns and 2 electrons (looking
apart from two separate atoms), in which the kerns are at rest''.
Furthermore, Sommerfeld \cite{Somm23} was clearly unaware of the key
asymmetric configuration of Fig. \ref{f1c}.
\bibitem{dot}  Throughout this paper ``exact'' energies (solid dots) are
taken from a combination of experimental and theoretical sources. The
theoretical analyses are typically many term (10-100) configuration
interaction computer calculations, see, for example, Ref. \cite{Scha84}.
\end{thebibliography}
\end{document}